\documentclass[english,aps,prl,amsmath,amssymb,twocolumn,letterpaper,citeautoscript,superscriptaddress,longbibliography]{revtex4-1}
\usepackage{mathtools} %floor and ceiling functions

\usepackage[normalem]{ulem} %Strikethrough
\usepackage{comment}
\usepackage{amsmath}
\usepackage{physics}
\usepackage{graphicx}
\usepackage{babel}
\usepackage{amstext}
\usepackage{color}
\usepackage{esint}
\usepackage[bookmarks=false,linkcolor=blue,urlcolor=blue,colorlinks,citecolor=blue]{hyperref}
\usepackage{pdfpages}	% Packages needed
\usepackage{pgffor}		% to include supplement

\makeatletter
\AtBeginDocument{\let\LS@rot\@undefined} %This fixes the issue
\makeatother							 %that pdfpages rotates the pages by 90deg.
\begin{document}

\title{Optical absorption tensors based on C$_{70}$ trimers and polymers} 

 \author{Elnaz Rostampour}
 \affiliation{Department of Physics, Urmia University, Urmia, Iran }
 \author{Badie Ghavami}
 \affiliation{School of Nano Science, Institute for Research in Fundamental Sciences (IPM), Tehran 19395-5531, Iran}
\author{Karin Larsson}
\affiliation{Department of Chemistry-Angstrom Laboratury, Uppsala University, Sweden }
\date{\today}
\begin{abstract}
The optical absorption spectrum of $C_{60}$-dimers and polymers was investigated by Kikuo et al. in 1996\cite{harigaya1996charge}.
As a compliment to these earlier studies, the optical absorption spectrum of the $C_{70}$ fullerene has been investigated in the present study. The main purpose was then to compare the absorption spectrum of the $C_{70}$-dimers and trimers and, more specifically, to clarify the effect of these molecular structures on the absorption spectrum. What is most important and decisive is then the value of the conjugation parameter of these $C_{70}$-based molecules. 
In the present study, a tight-binding model was used in calculating the optical absorption spectra of both $C_{70}$ dimers and polymers, as
well as $C_{70}$ trimers and polymers. The change  in  conjugation parameter for each of these species  was found to cause  variations in the corresponding optical absorption spectrum.  It was found that the absorption tensor of the $C_{70}$ trimer and the polymer was depending on the value of the conjugation parameters $b=0.5$  and $b=0.8$. The situation was almost the same for the conjugation parameters $b=0.1$ and $b=0.2$. In addition, the value of the band gap was also different depending on the different conjugation parameters, with a reduced value for the larger values of this parameter.
As a conclusion, smaller values of the conjugation parameter was not found to have a large effect on the absorption spectrum of the $C_{70}$-dimers and trimers, or in other words, the effect was hardly visible. On the contrary, the larger values caused a drastic change in the optical absorption spectrum of the $C_{70}$-dimers and trimers.
\end{abstract}
\maketitle

\section{Introduction}

It has recently been found that the polymerization of $C_{70}$ fullerenes takes place by a spatial regulation of the reactive double
bonds in neighboring cages so that the alignment of cages,  and  their  hypothetical directional mobility, supports the
polymerization.  All observed  dry  polymerization  events  have been found to be clearly attributed  to  the  results  of  topochemistry\cite{sheka2011fullerene}.
Nanocomposites of fullerene-based  polymers  have  many practical applications, all the way from solar  cells\cite{huang2010polypropylene} to new materials\cite{badamshina2008characteristics,karpacheva2000fullerene,wang2004zhu}. 
Examples include  optical  interrupters and photodiodes, photo-conductors, and electrodes for lithium batteries\cite{badamshina2008characteristics}.
Polymeric fullerene  materials  can  be  constructed  by  many methods.  For  example,  it can be constructed by  irradiation  with  
electrons or ions, treatment  in  a  plasma  generator,  doping  with  alkali metals\cite{dresselhaus1996science,renker1999alignment,stephens1994polymeric}, direct chemical synthesis\cite{wang1997synthesis}, or mechanical milling\cite{liu1999structural}.
With the  $N_2$   laser  that  is  used  to  measure  the mass  spectrum  of  laser  desorption,  the  irradiation  of  $C_{70}$ dimers or $C_{70}$ trimers in the 
mass spectrometer causes the initiation of photopolymerization\cite{rao1993photoinduced,cornett1993laser}.
 Several  experimental and theoretical investigations on  (C$_{60}$)$_2$, (C$_{70}$)$_2$ and even (C$_{59}$N)$_2$ have  been reported\cite{fujitsuka1999triplet,lebedkin2000structure,heine2001isomers,hummelen1995isolation}.
It is, thus, important to elucidate  the  unique  physical  properties of  
fullerene  polymeric materials.
 The primary experimental synthesis and characterizations of five stable fullerene dimers $(C_{70})_2$ with [2+2] bridges between hexagon-hexagon bonds carried out by S. Forman  et al.\cite{forman2002novel}\\
 One  of  the  attractions  with the  $C_{60}$  dimers  and  trimers is the evaporation of $C_2$ units. The average number of 
evaporated $C_2$ units  from  the  $C_{60}$ dimer,  trimer,  tetramer, and  pentamer  in  the  mass  spectra specify which type of characterization method that was used for this analysis has been  obtained  by integrating 
the mass peaks. Considering the radical anion spectroscopic signatures of the photogenerated polymers,  $C_{70}$-fullerene 
It is  $C_{70}$ acoording to ref.\cite{liedtke2010spectroscopic} derivatives and dimers mixed with different polymers was investigated  by  Liedtke  et  al.
By using photo-induced  absorption  (PIA) and  electron  spin  resonance  spectroscopy  (ERS),  common  characteristics  of $C_{70}$ have  been found  to  be  an  additional  PIA  peak  below  the band gap 
below the conduction band minimum of about $0.92$ eV and an ESR shoulder around $g \geq 2.005$.  
A comparison  of results  from  several  compounds  showed 
that  this  characteristic  is  the  signature  of  the $C_{70}$ radical  anion. 
Thus, the side chains of the fullerenes or the charge transfer states at polymer:C70-fullerene interfaces have no role in C70 spectral features in PIA and ESR spectroscopy.   
Low-frequency ESR  features  and  an additional photoinduced  absorption  peak  have also been found to be present in  $C_{70}$ - based  hetero- and  homodimers\cite{delgado2009fullerene}.
Furthermore, the  interacting  electron-phonon  model  was  used  by  Harigaya\cite{harigaya1995metal,harigaya1996doping}  in studying the electronic 
structure of one-dimensional $C_{60}$ polymers. 
 In  the  direction  of  the  polymer  chain,  the strength  of  the electron  conjugations was indicated  by  a  phenomenological  parameter. A strong  conjugation  corresponds to a large  charge transfer  integral  between  the $C_{70}$ 
units in the polymer.  For a neutral polymer\cite{harigaya1995metal}
with a strong conjugation, a  level  crossing  can been  formed between  the  highest  
occupied electronic state and the  lowest  unoccupied electronic state.
Moreover, electron doping (e.g., doping with metals) has been found to have a large effect on the properties of $C_{60}$ polymers.
 They presented metallic properties when doping with one electron per $C_{70}$ unit. Furthermore, a strengthening of  the  electron conjugation took place when doping with two electrons per $C_{70}$ unit. As a result,  the $C_{70}$ polymer became an insulator, and it went from an indirect bandgap to a direct bandgap
material\cite{harigaya1996charge}.
In addition to  the  experimental  research activities, the optical properties of $C_{70}$ fullerenes\cite{harigaya1994optical} and solids\cite{harigaya1994exciton, deshpande199522nd} have also been theoretically analyzed. 
The main goal with the present study has been to study the optical absorption spectra of  the  polymerized  $C_{70}$ systems using  theoretical tight-binding  (TB)  calculations.
More specifically, the (C$_{70}$)$_2$, (C$_{70}$)$_3$, and their respective polymer systems  have then been investigated. 
%%%%%%%%%%%%%%%%%%%%%%%%%%%%%%%%%%%%%%%%%%%%%%%%%
\section{Theoretical model}
The tight-binding model for the C$_{70}$ polymer backbone can be written as
\begin{widetext}
\begin{equation}
	\label{q1}
	\begin{split}
H=-bt\sum_{l,\sigma}\sum_{<i,j>=<1,3>,<2,4>}(c^\dagger_{l,i,\sigma}c_{l+1,j,\sigma}+h.c.)\\
&
-(1-b)t\sum_{l,\sigma}\sum_{<i,j>=<1,2>,<3,4>}(c^\dagger_{l,i,\sigma}c_{l,j,\sigma}+h.c.)\\
-t\sum_{l,\sigma}\sum_{<i,j>=others}(c^\dagger_{l,i,\sigma}c_{l,j,\sigma}+h.c.)
 \end{split}
\end{equation}
\end{widetext}
where t is the charge transfer integral between nearest-neighbour carbon atoms.
l is the lth $C_{70}$ molecule and $<i,j>$ is the pair of neighbouring ith and jth atoms.
Also, $c_{l,i,\sigma}$ is the annihilation operator of the electron with spin $\sigma$ 
at the ith site of the lth molecule. Finally, is the sum is over the pairs of neighbouring atoms.
\begin{figure}[htp]
    \centering
	\includegraphics[scale=0.17]{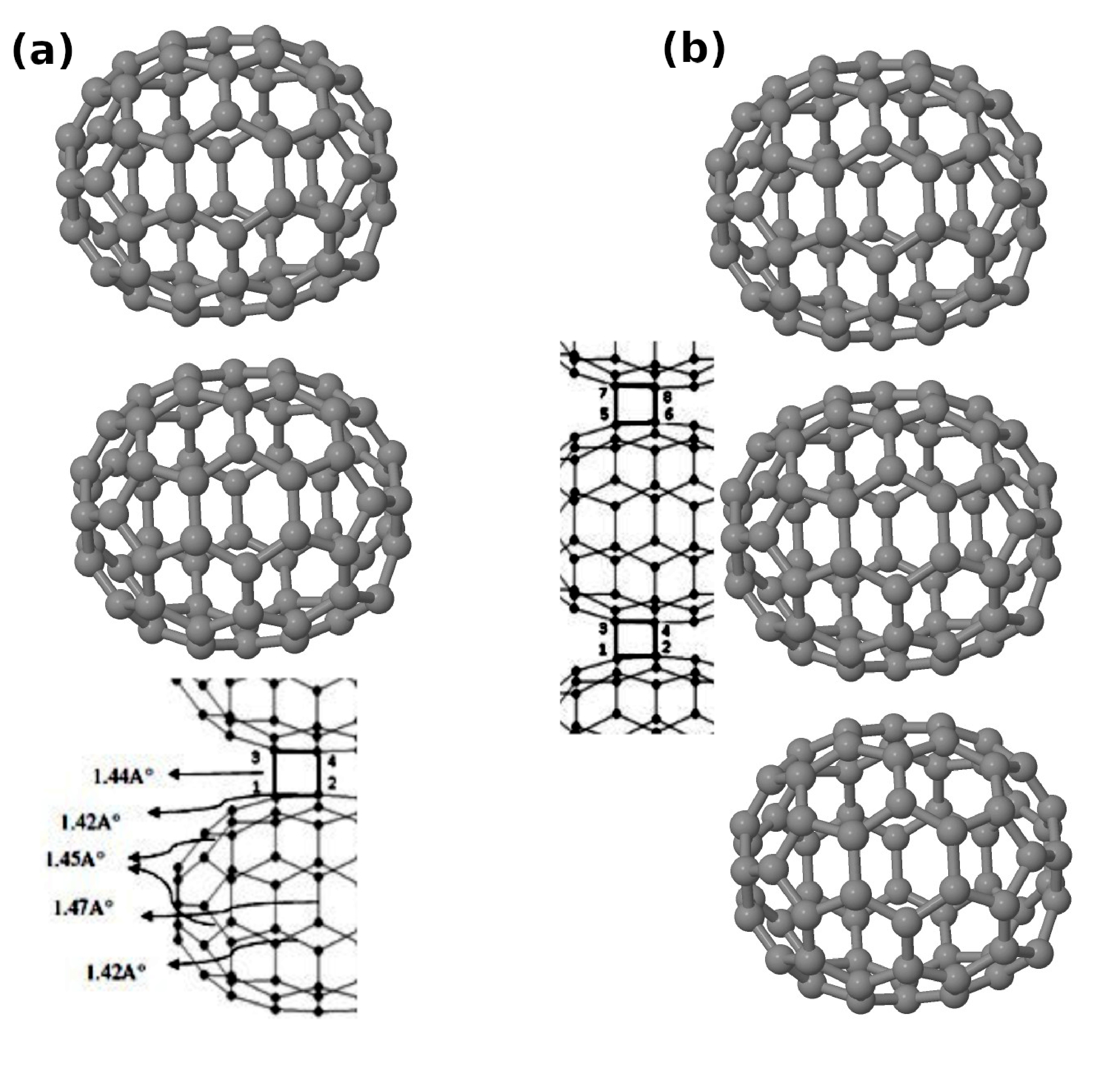}
		\caption{Structures of (a) two and (b) three $C_{70}$ unit polymer systems. The carbon sites which constitute the four-membered rings are marked  with numbers. The bond lengths that are used in the calculations are also presented.}
		\label{fig1}
\end{figure}
As can be seen in Fig.\ref{fig1}, the atoms with $i = 1-4$ in the four-membered ring are denoted by numbers. Among the neighbouring fullerene molecules, the strength of the conjugations can be controlled by a parameter b Ref\cite{harigaya1995metal,harigaya1996doping}.
For $b = 1$, the $\sigma$ bond between atoms with numbers 1 and 2 has disappeared. 
This is also the case with the $\sigma$ bond between atoms with numbers 3 and 4. In these cases, the orbitals are similar to the $\pi$ orbitals.  On the other hand, there is a double bond between atoms 1 and 3, and this is also the situation for the atoms 2 and 4. 
By reducing the $b$ parameter, the conjugation of neighbouring molecules will be reduced, causing the $C_{70}$ molecules to act independently of each other.
Thus, the interaction between the fullerene molecules is smaller in the intermediate b region. The operator $c_{l,i}$, at the lattice sites of the four-membered rings indicated a molecular orbital.
The parameter b is varied within $0 \leq b\leq 1.0$.  The calculated value of t is 2eV. Two and three $C_{70}$ units in the polymer, with its bond lengths, are shown in Fig.\ref{fig1}. 
As can be seen in Fig.\ref{fig1}a, the bond lengths between the $C_{70}$ units (i.e., the 1-3 lengths and 2-4 bond lengths) are 1.44\AA. The other $C-C$ lengths within the $C_{70}$ units are 1.42, 1.45 and 1.47\AA. 
These lengths are obtained using the results of geometric optimization.\\
These lengths are the results from geometry optimizations. Depending on the direction of the applied electric field, the optical absorption spectra 
of $C_{60}$ was earlier found to be isotropic, while it was found to be anisotropic for $C_{70}$\cite{agranovich2013crystal}.

This finding can be explained  by the decreased symmetry in $C_{70}$ with respect to $C_{60}$. Anisotropy could, thus, be used in the present study to simplify the analysis of the calculated data.
$C_{70}$ dimers and polymers, as well as $C_{70}$ trimers and polymers. experience very large orientation disorders but the effect of anisotropy of the optical absorption spectrum is not experimentally observable. The sum-over-state (SOS) expression for the xy component of the polarizability tensor can be written as:
\begin{equation}
\label{q2}
	\alpha_{xy}=2\sum_m^{unocc}\sum_n^{occ}\frac{<\varphi_m|\hat{x}|\varphi_n><\varphi_m|\hat{y}|\varphi_n>}{|\epsilon_m -\epsilon_n|}
\end{equation}

where $\varphi_i$ is the orbital for electron i, and $\epsilon_i$ is the energy of $\varphi_i$. Furthermore, the average polarizability tensor 
$\alpha_{SOS}=\frac{1}{3}(\alpha_{xx}+\alpha_{yy}+\alpha_{zz})$ was calculated according to Eq.\ref{q2} once the self-consistent response of the Kohen and Sham equations was obtained for each fullerene unit\cite{ivanov2001photoionization,amusia2004random}. The optical absorption tensor was, thereafter, calculated using Eq.\ref{q3}: 
\begin{equation}
\label{q3}
\sigma(\omega)=\frac{4\pi\omega}{c}Im\left[\alpha\left(\omega\right)\right]. 
\end{equation}
%%%%%%%%%%%%%%%%%%%%%%%%%%%%%%%%%%%%%%%%%%%%%%%%%%%%%
\section{Results and discussions}
High-temperature photopolymers of $C_{60}$ units are believed to mainly consist of  dimers $(C_{60})_2$, but larger polymers 
are formed at room temperature. Optical absorption spectroscopy has indicated that each fullerene unit is bound to 1-3 other units\cite{waagberg1999comparative,pusztai1999bulk}, 
and the resulting polymer structures have the form of chains, branched chains\cite{waagberg1999comparative}, or rings with three 
or more fullerene units\cite{pusztai1999bulk}. Photopolymerization of $C_{70}$ has also been found to be possible. However, since only five $C=C$ bonds have a reasonable reactivity near the poles of $C_{70}$, the polymer yield is smaller than for $C_{60}$ and the polymer size seems limited to mainly dimers, $(C_{70})_2$, and trimers, $(C_{70})_3$\cite{menon1995anisotropic}.
Calculations have in the present study been performed for the $C_{70}$-based polymers. The $(C_{70})_2$ dimers and $(C_{70})_3$ trimers are shown in  Fig.\ref{fig1}. 
More specifically, numerical calculations has been performed for the $C_{70}$-based dimers, trimers, and respective polymers by using tight-binding methods.
The $xx$, $yy$, and $zz$ elements of the absorption tensor, in addition to the average elements, $\frac{1}{3}(\sigma_{xx}+\sigma_{yy}+\sigma_{zz})$, for different b-parameters, where then calculated for the model systems. The calculations showed that the $\sigma_{xx}$, $\sigma_{yy}$, and $\sigma_{zz}$ spectra were not identical for the $C_{70}$ dimers and trimers. The orientation of the $C_{70}$ unit relative the applied electric field of light played a decisive role for the determination of the elements in the optical absorption tensor. \\
 Also, the anisotropy of the absorption tensor was due to the decrease in symmetry of the $C_{70}$ unit. For polymer systems, the possibility for crossover of electronic states can take place for Frenkel excitons and Wannier excitons. The excitons are located on the molecule in the absence of interactions between molecular units in a polymer. However, in the presence of intermolecular interactions, the amplitude of the excitations (such as charge-transfer excitons) are dependent 
on the number of molecular units in the polymer. 
\begin{figure}[htp]
    \centering
	\includegraphics[scale=0.5]{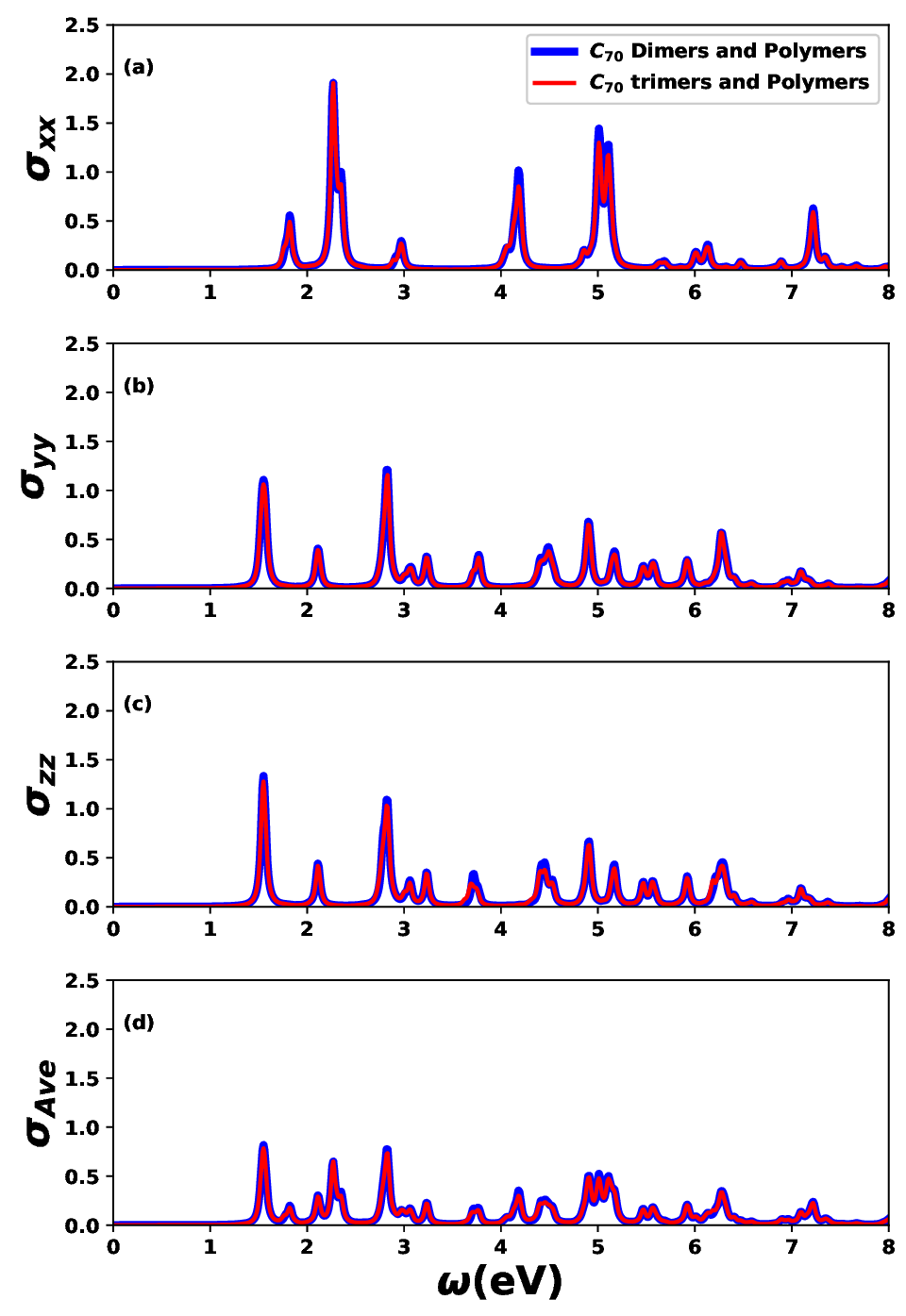}
		\caption{The (a) xx, (b) yy, (c) zz elements of the optical-absorption tensor, and (d) average of the optical-absorption tensor of $C_{70}$ dimers and polymers, as well as $C_{70}$ trimers and polymers with $b=0.1$. }
		\label{fig2}
\end{figure}
\begin{figure}[htp]
    \centering
	\includegraphics[scale=0.5]{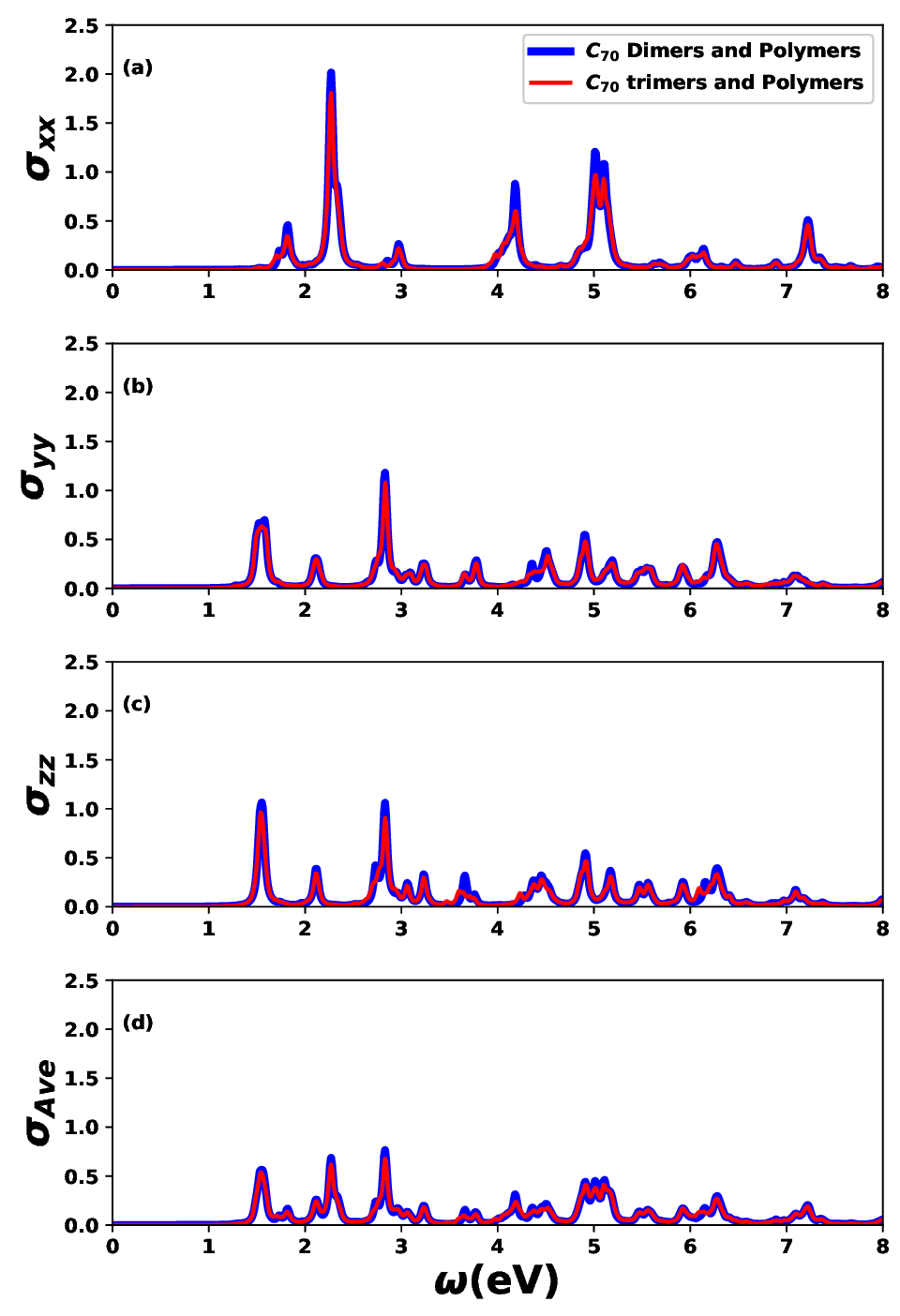}
		\caption{For $C_{70}$ dimers, trimers, and respective polymers with $b=0.2$ (a) $xx$, (b) $yy$, and (c) $zz$ elements, as well as (d) averages, of the optical absorption tensors. }
		\label{fig3}
\end{figure}
Fig.\ref{fig2} and Fig.\ref{fig3} show the absorption tensors of the$C_{70}$ dimers and trimers, in addition to their respective polymers, for weaker conjugations, i.e., $b=0.1$ and $0.2$. It was found that the absorption Fig.\ref{fig2}.\\
For $C_{70}$ dimers, trimers, and respective polymers with $b=0.1$ and $0.2$; (a) $xx$, (b) $yy$, and (c) $zz$ elements, as well as (d) averages, of the optical absorption tensors.
tensor of the $C_{70}$ dimer was almost identical to the absorption tensor of the corresponding polymer (i.e., with a $C_{70}$ dimer as the unit in the polymer). 
This was also the situation with the $C_{70}$ trimer and its corresponding polymer (i.e., with a $C_{70}$ trimer as the unit in the polymer. However, the dipole transitions in the energy region below $1.55$ eV were prohibited for the single $C_{70}$ unit. Fig.\ref{fig4}, and Fig.\ref{fig5} show the absorption tensors of  the $C_{70}$ dimers, trimers, and their respective polymers for stronger conjugations, i.e., $b= 0.5$ and $0.8$. The absorption tensors were different for this situation, and this was also the situation with the corresponding optical absorption spectra.
There were more peaks in the absorption spectra (i.e. for both $b=0.5$ and $0.8$) for stronger interunit interactions.
As can be seen in Fig.\ref{fig4}d, there was also a small shoulder in the energy region below $1.08$ eV. This region overlaps with the forbidden dipole transition for the single $C_{70}$ unit. However, these transitions are allowed in case of intermolecular interactions, which can be explained by the broad structure of the absorption spectrum at low energies and $b=0.5$, and $0.8$.\\
Similar $C_{60}$ polymer structures have been synthesized at high pressures, and characterized by optical absorption\cite{ogawa1998optical}. 

\begin{figure}[htp]
    \centering
	\includegraphics[scale=0.48]{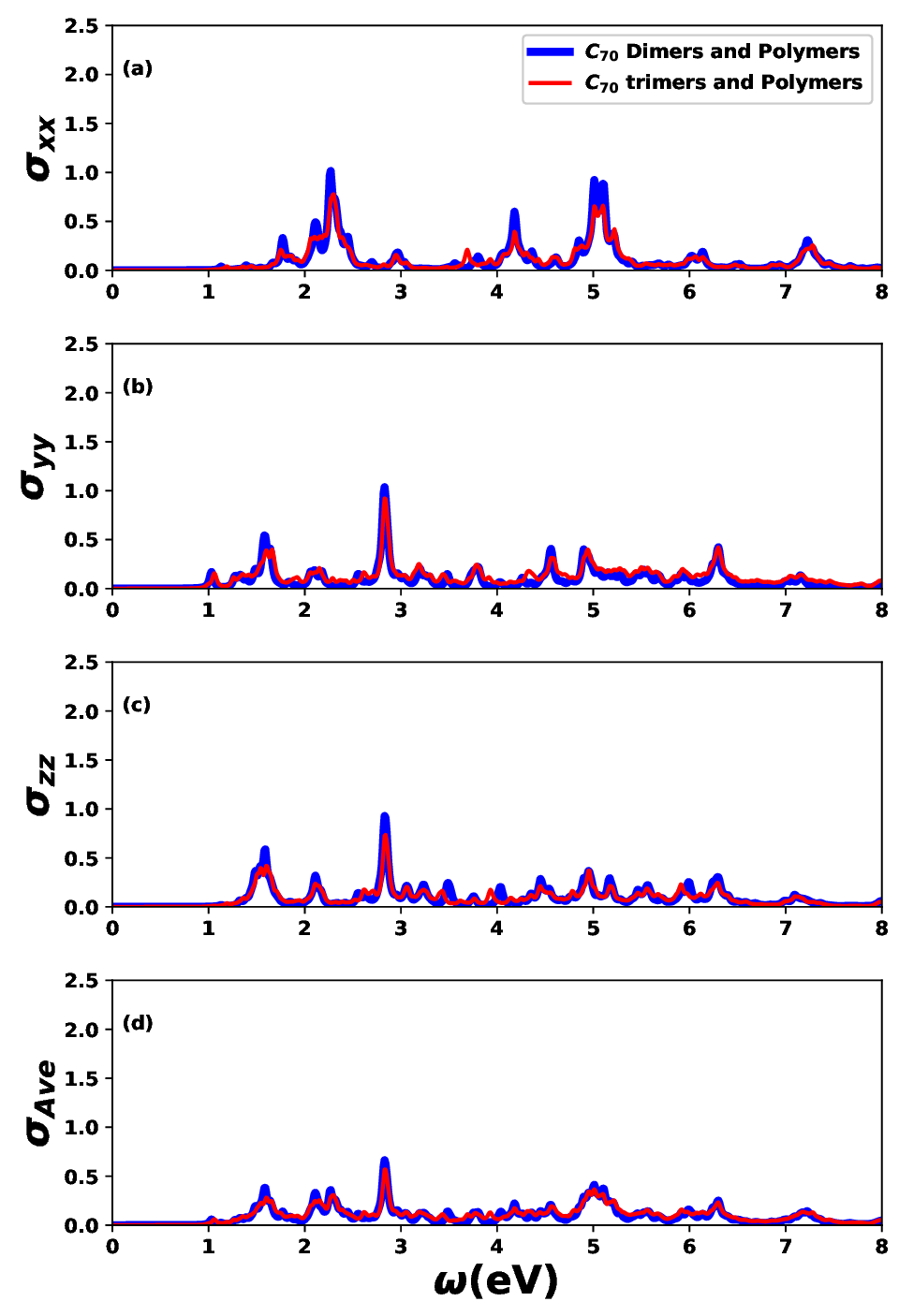}
		\caption{For $C_{70}$ dimers, trimers, and  respectivepolymers with $b=0.5$ (a) $xx$, (b) $yy$, and (c) $zz$ elements, as well as (d) averages, of the optical absorption tensors. }
		\label{fig4}
\end{figure}
\begin{figure}[htp]
    \centering
	\includegraphics[scale=0.48]{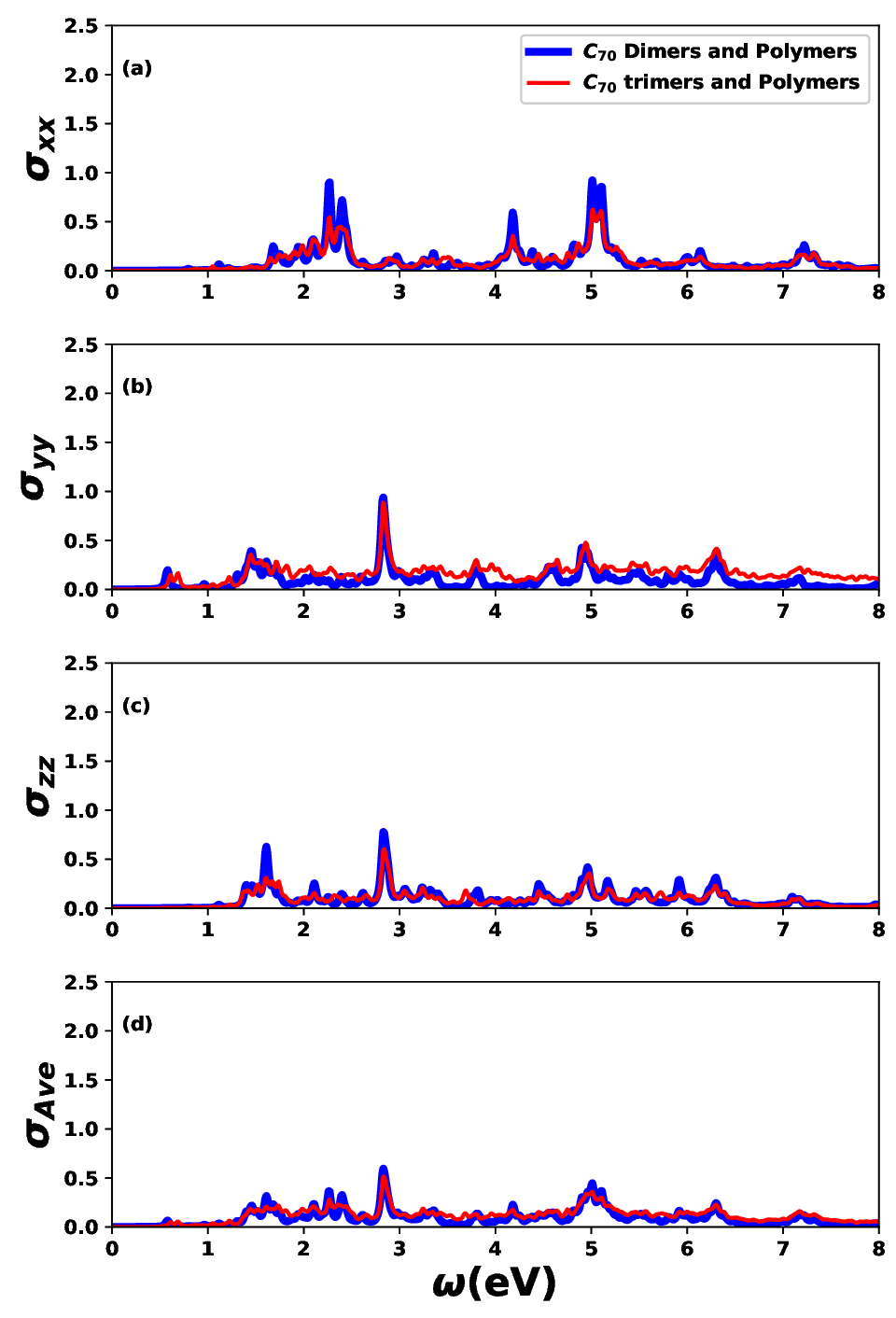}
		\caption{For $C_{70}$ dimers, trimers, and respective polymers with $b=0.8$ (a) $xx$, (b) $yy$, and (c) $zz$ elements, as well as (d) averages, of the optical absorption tensors. }
		\label{fig5}
\end{figure}

 It is known as the symmetry reduction effect. As can be seen in Fig.\ref{fig5}d, the first peak in the optical absorption tensor is centered at $\omega = 0.59$ eV for the $C_{70}$ dimers and polymers. Furthermore, the fluctuations in the absorption spectrum were most prominent for $b=0.8$\\

 \begin{figure}[htp]
    \centering
	\includegraphics[scale=0.4]{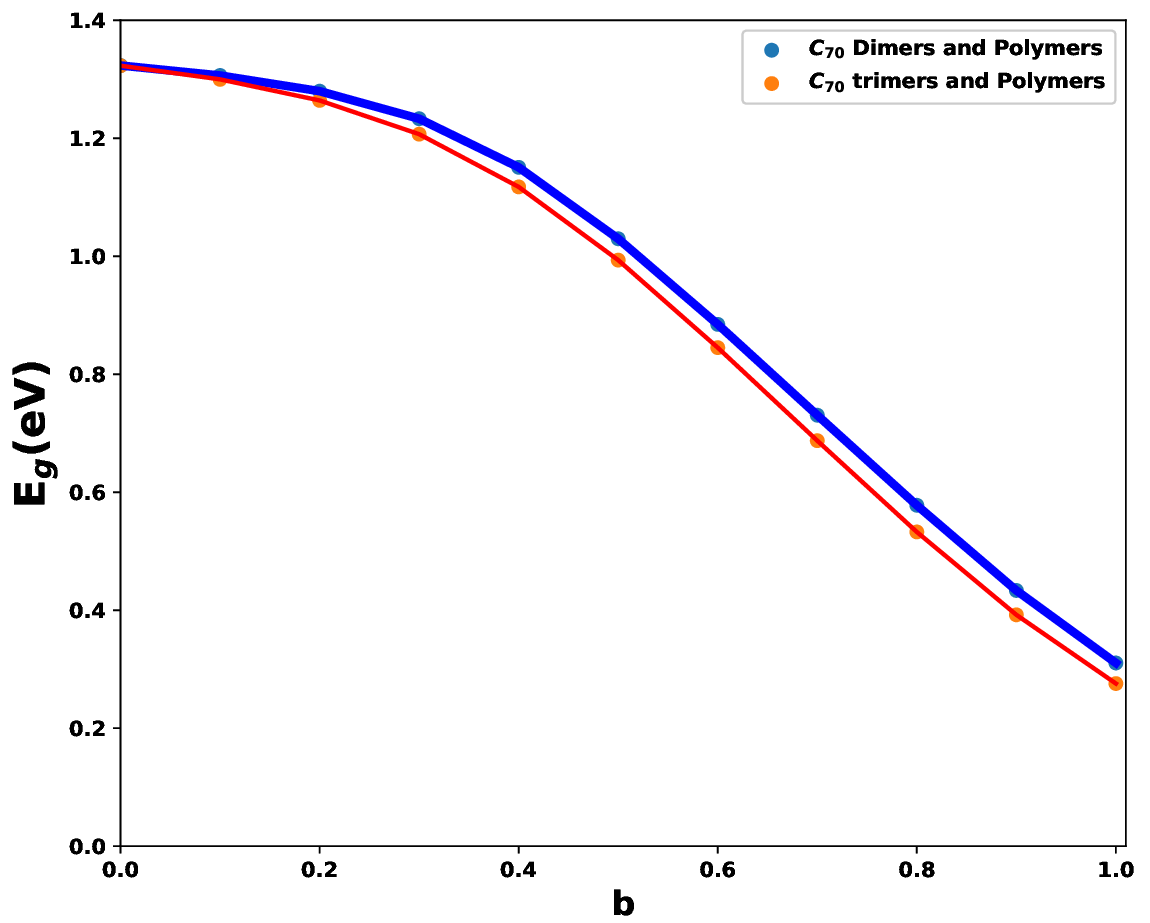}
		\caption{Band gap values of the C70 dimers, trimers, and respective polymers as a function of the parameter b. }
		\label{fig6}
\end{figure}

The band gap values for the $C_{70}$ dimers, trimers, and respective polymers in terms of functions of the parameter b have also been calculated. As can be seen in Fig.\ref{fig6}, the band gap value decreased for an increased value of b. For carbon systems with $sp^2$ conjugations, the optical conductivity peaks are shown in the energy regions of the transitions between 
the $(\pi/ \sigma)$ and the $(\pi^*/ \sigma^*)$ electronic states. Due to the high symmetry of the $C_{60}$ molecule, and the weak van-der-Waal's interactions in between\cite{sohmen1993electron}, these peaks became very distinct. Furthermore, three broad features were observed for graphite at the energies $4.56 \pm 0.05$ eV, $13 \pm 0.05$  eV, and $15 \pm 0.05$ eV, and for single-wall carbon nanotubes at $4.3 \pm 0.1$ eV, $11.7 \pm 0.2$  eV, and $14.6 \pm 0.1$ eV. The optical conductivity had also additional structures at a low energy for the single-wall carbon nanotubes\cite{rostampour2021effect}.\\

\section{Concluding remarks }

The effects of the conjugation parameter, b, on the absorption tensor of the $C_{70}$ trimers, dimers, and their respective polymers have been investigated in the present study. It was found that the optical spectra and the bandgap values 
changed by changing b. The conjugations were also observed to affect the charge transfer integrals between the $C_{70}$ molecular units, so that a stronger conjugation corresponded to a larger charge transfer integral. More precisely, the conjugation between molecular units decreased with a decrease in b. In other words, the intermolecular interactions in the region between these two molecular units decreased. For the situation with b=0, it meant that the two molecular units 
were completely independent of each other and there was no charge transfer in between. Considering the highest occupied electronic state and the lowest unoccupied electronic state in the polymers, the level crossing between these two states occurred for an increasing degree of conjugation in the neutral polymer.

%\bibliographystyle{apsrev4-1}
%\bibliography{refs}

\end{document}